\documentclass[11pt,draftclsnofoot,onecolumn]{IEEEtran}

\usepackage[english]{babel}

\usepackage{amssymb}
\usepackage{amsmath}
\usepackage{cite}
\usepackage{url}
\usepackage{xcolor}
\usepackage{cite,graphicx,amsmath,amssymb}
\usepackage{fancyhdr}
\usepackage{mdwmath}
\usepackage{mdwtab}
\usepackage{caption}
\usepackage{amsthm}
\usepackage{setspace}
\usepackage{algorithm}
\usepackage{algorithmic}
\usepackage{pdfpages}
\usepackage{mathtools}
\usepackage{subcaption}
\usepackage{makecell}
\usepackage[most]{tcolorbox}
\usepackage{mdframed}
\usepackage{cuted}
\usepackage{bm}

\DeclareMathOperator*{\argmax}{argmax}
\DeclareMathOperator*{\argmin}{argmin}

\DeclareFontFamily{U} {MnSymbolA}{}
\DeclareFontShape{U}{MnSymbolA}{m}{n}{
  <-6> MnSymbolA5
  <6-7> MnSymbolA6
  <7-8> MnSymbolA7
  <8-9> MnSymbolA8
  <9-10> MnSymbolA9
  <10-12> MnSymbolA10
  <12-> MnSymbolA12}{}
\DeclareFontShape{U}{MnSymbolA}{b}{n}{
  <-6> MnSymbolA-Bold5
  <6-7> MnSymbolA-Bold6
  <7-8> MnSymbolA-Bold7
  <8-9> MnSymbolA-Bold8
  <9-10> MnSymbolA-Bold9
  <10-12> MnSymbolA-Bold10
  <12-> MnSymbolA-Bold12}{}

\DeclareSymbolFont{MnSyA} {U} {MnSymbolA}{m}{n}

\tcbuselibrary{breakable} 

\title{\huge Near-Field Localization and Sensing with Large-Aperture Arrays: From Signal Modeling to Processing}

\author{\IEEEauthorblockN{Zhaolin Wang, Parisa Ramezani, Yuanwei Liu, and Emil Bj\"ornson}
\thanks{Zhaolin Wang is with the School of Electronic Engineering and Computer Science, Queen Mary University of London, London, U.K. (email: zhaolin.wang@qmul.ac.uk). Parisa Ramezani and Emil Bj\"ornson are with KTH Royal Institute of Technology, Stockholm, Sweden (email:\{parram,emilbjo\}@kth.se). Yuanwei Liu is with the Department of Electrical and Electronic Engineering, The University of Hong Kong, Hong Kong SAR (email: yuanwei@hku.hk).} 
\vspace{-1cm}
}

\begin{document}

\maketitle

The signal processing community is currently witnessing a growing interest in near-field signal processing, driven by the trend towards the use of large aperture arrays with high spatial resolution in the fields of communication, localization, sensing, imaging, etc. From the perspective of localization and sensing, this trend breaks the basic far-field assumptions that have dominated the array signal processing research in the past, presenting new challenges and promising opportunities.

\section{Introduction}
Localization and sensing (L\&S) are critical functionalities in modern wireless systems, particularly with the advent of advanced multiantenna technologies. Localization refers to the process of acquiring the states, such as location and velocity, of connected devices within wireless networks. Tracing back to the inception of first-generation (1G) standards, this capability has been historically viewed as an optional function of wireless cellular networks \cite{ott1977vehicle}. Over the decades, various localization methods have been developed which include terrestrial techniques that utilize time of arrival (TOA), time-difference-of-arrival (TDOA), or angle of arrival (AOA) measurements, and satellite-based approaches represented by the Global Positioning System (GPS). These approaches rely on sending signals from a source to a receiver, and processing the received signal to extract relevant state information. Sensing, on the other hand, refers to the estimation of the states of passive, non-cooperative reflecting objects, similar to traditional radar systems. Sensing is expected to become a native function in next-generation wireless networks because it enables new services and is made possible since base stations will have  similar multiantenna and spectrum features as digital radars. 

L\&S techniques have advanced with multiantenna technologies, evolving from phased arrays to digital antenna arrays, and continuously moving towards larger multiantenna systems. The concept of using multiple antennas, or arrays, for enhancing target localization and detection dates back to the early 20th century, with significant advancements during World War II for radar applications. These early developments laid the groundwork for understanding how antenna arrays could be manipulated to improve directionality and gain, which are crucial for L\&S. The development of phased-array antennas for radar in the 1950s was a significant milestone \cite{fenn2000development}. In this era, a new initiative is to steer the direction of the beam electronically without moving the physical antennas, thereby greatly enhancing the speed and flexibility of L\&S. 
With the advancement of antenna array technologies, array signal processing gradually became an active research area since the 1980s \cite{krim1996two}. This period, extending into the 1990s, witnessed a surge of classical L\&S methods based on array signal processing, including the famous MUSIC (MUltiple SIgnal Classification) method, parametric MLE (Maximum-Likelihood Estimation) method, and ESPRIT (Estimation of Signal Parameters via Rotational Invariance Technique) method. These methods laid an important foundation for array signal processing techniques for L\&S. The turn of the century brought about a digital revolution, further transforming the capabilities of antenna arrays. The integration of digital signal processing with antenna array technology led to the development of adaptive antenna systems capable of dynamically adjusting their patterns to optimize L\&S accuracy. This era also witnessed the advent of Multiple-Input Multiple-Output (MIMO) systems \cite{bjornson2023twenty}, which use multiple antennas at both the transmitter and receiver to improve communication performance through beamforming, spatial multiplexing, and diversity. Building on these advancements, the concept of MIMO radar emerged in the early 2000s \cite{1316398}, allowing each antenna to transmit different waveforms and thus providing more adaptive array configuration and higher L\&S resolution.

The electromagnetic (EM) radiation field emitted by antennas or antenna arrays can be divided into two regions: the far-field (Fraunhofer) region and the radiative near-field (Fresnel) region \cite{Ramezani2023a, 10220205}. The demarcation between these regions is traditionally defined by the phase error-based Fraunhofer distance (also known as the Rayleigh distance), which is given by 
\begin{equation}
    d_{\mathrm{FA}} = \frac{2D^2}{\lambda},
\end{equation}
where $D$ denotes the aperture of the antenna (array) and $\lambda$ is the signal wavelength. The exact boundary between these regions can be defined differently depending on the application. Regardless, when the emitted signal propagates into the far-field region, its wavefront is approximated as planar, which implies that the receiving antenna array can only resolve the direction of the signal. However, the accurate spherical-wavefront model should be used to characterize the signal propagation within the near-field region, where the curvature is distance-dependent. This property allows the antenna array to determine both the direction and the propagation distance of the signal.
Research on L\&S of sources or targets in the near-field region dates back to as early as the 1980s \cite{rockah1987array, Huang1991Near}. However, this research topic did not attract widespread attention during that era. This is because the array aperture and the signal frequency were generally limited, rendering the sources or targets normally located in the far-field region.

The landscape shifted with the introduction of the massive MIMO concept by Marzetta in 2010 \cite{5595728}. The core idea of massive MIMO is to equip base stations with numerous antennas connected to fully digital transceiver branches, allowing them to generate many simultaneous narrow beams with high controllability. Massive MIMO was developed for communications but builds on enhancing the spatial resolution compared to smaller arrays, which also significantly improves L\&S accuracy. These favorable features have made an elementary form of massive MIMO with 32-64 antennas the fundamental building block of 5G networks, typically operating in the 3.5 GHz band.
Since the emergence of 5G massive MIMO, researchers have continued to envision using even larger arrays with various form factors and characteristics, inevitably increasing the physical size of the antenna arrays.
Coupled with the trend towards exploiting high-frequency bands (centimeter-, millimeter-waves, and beyond), which have very short wavelengths, the near-field region around base stations can be significantly expanded.  For instance, an antenna array with a 1-meter aperture operating at 30 GHz can have a near-field region extending up to 200 meters, making it highly likely for targets to be located within this near-field region. Against the above background, a lot of research efforts have been devoted into the field of near-field L\&S, focusing on various aspects such as signal modeling \cite{Ramezani2023a, 10220205, 8736783, 9139337}, signal processing \cite{Zhi2007nearfield, He2012efficient, Zhang2018reduced, cui2022channel, 9762062, wang2023velocity, Huang2024low}, and performance bound characterization \cite{el2010conditional, giovannetti2024performance}. This article aims to provide a tutorial review on near-field L\&S from signal modeling to signal processing. Our goal is to highlight the most significant changes from traditional far-field L\&S and to elucidate the key intuitive ideas for leveraging the advantages of near-field L\&S and addressing the associated challenges. 


\section{Near-Field Properties}
\begin{figure}
	\centering
	\begin{subfigure}{0.45\columnwidth}
		\centering
		\includegraphics[width=1\columnwidth]{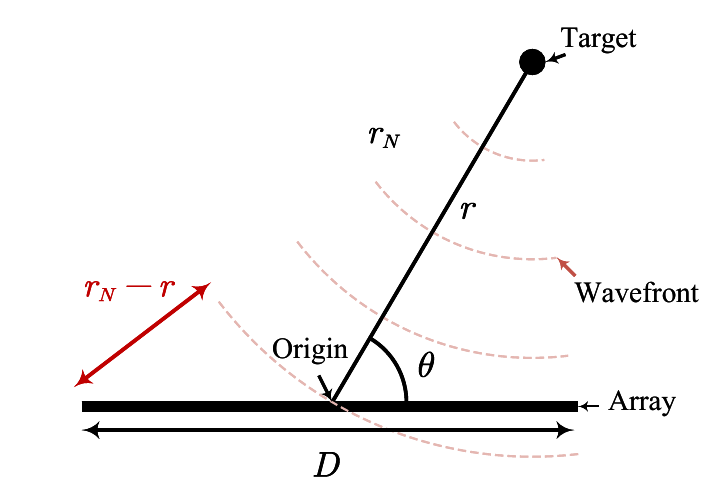}
		\caption{Near-field model}
	\end{subfigure}
	\begin{subfigure}{0.45\columnwidth}
		\centering
		\includegraphics[width=1\columnwidth]{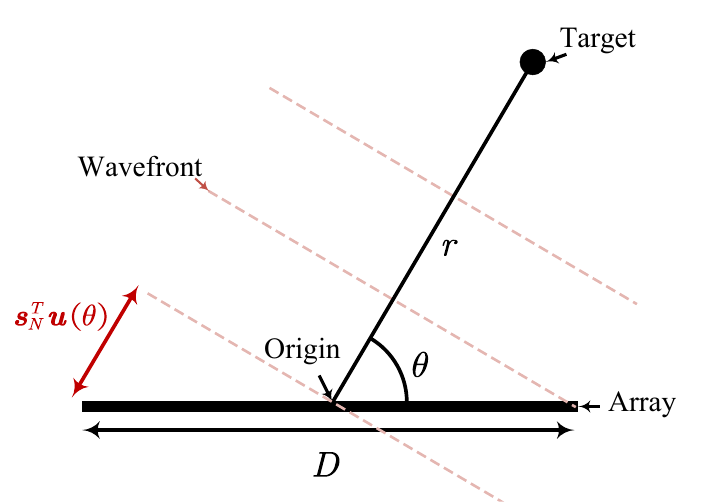}
		\caption{Far-field model}
	\end{subfigure}
 \caption{The schematic of the near-field and far-field models under spherical wavefront and planner wavefront assumptions, respectively.}
 \label{fig:model}
  \vspace{-0.7cm}
\end{figure}

In this section, we will first outline near-field propagation from the perspective of EM theory and discuss its relevance to the near-field L\&S considered in this article. Let us consider a sinusoidal wave with frequency $f_c$ transmitted from a point source. The Fourier representation of the electric field observed at a distance $r$ from the source, in any direction perpendicular to the propagation direction, is proportional to \cite{9139337}
\begin{equation}
    G(r) = - \frac{j \eta e^{- j \frac{2\pi}{\lambda} r}}{2 \lambda r} \left( 1 +  \frac{j \lambda }{2 \pi r} 
    - \frac{\lambda^2}{(2 \pi r)^2}   \right),
\end{equation}
where $\eta$ denotes the intrinsic impedance, $\lambda = c/f_c$ denotes the wavelength with $c$ being the speed of light, and $j = \sqrt{-1}$. The last two terms, which decay faster than $1/r$, diminish rapidly as $r$ increases and do not contribute to EM radiation. These terms are associated with “near-field” propagation in EM theory, but they are more commonly referred to as “reactive near-field” in recent literature on near-field communications and L\&S. It has been demonstrated that the reactive near-field is confined within a few wavelengths from the source, even for an infinitely large antenna (or array) \cite{10614327}. Therefore, in practical system design, only the radiating field needs to be considered, where the electric field at a distance $r$ is proportional to
\begin{equation}
    G(r) = - \frac{j \eta e^{- j \frac{2\pi}{\lambda} r}}{2 \lambda r}.
\end{equation}
In the radiating field, the electric field has an amplitude proportional to $1/r$ and a phase of $2 \pi r / \lambda$. According to conventional EM theory, the entire radiating field from a single small antenna is defined as the “far-field” \cite{8736783}. However, recent literature on near-field communications and L\&S using antenna arrays further divides this field into “radiating near-field” and “radiating far-field,” based on whether the wavefront of waves emitted from a large radiating aperture or antenna array is spherical or planar, as illustrated in Figure \ref{fig:model}. The terms “near-field” and “far-field” discussed in this article refer specifically to the “radiating near-field” and “radiating far-field” rather than their classical definitions in EM theory. The reactive near-field is omitted due to its negligible impact.

\section{End-to-end signal modeling}
For L\&S, the location and velocity of targets are of the most interest to acquire. The method of obtaining these parameters depends highly on the channel structure. In this section, we step through the end-to-end signal modeling of near-field L\&S from fixed to moving targets, and reveal how near-field effects influence the way antenna arrays observe target parameters. We focus primarily on the case involving a single L\&S node. Distributed L\&S systems with multiple nodes are beyond the scope of this article.

\subsection{Fixed targets}

We begin with signal modeling for the L\&S system with a fixed target, using a two-dimensional coordinate system and a point-like target for simplicity. As shown in Figure \ref{fig:model}, we assume the L\&S node, equipped with an $N$-antenna large-aperture array, is centered at the origin with its $n$-th antenna located at $\mathbf{s}_n \in \mathbb{R}^2$. The aperture of the array is defined as $D = \max_{m,n} \| \mathbf{s}_n - \mathbf{s}_m \|$, where $\|\cdot\|$ is the Euclidean norm. The target's location is denoted by $\bm{p} = r \bm{u}(\theta)$, where $r$ is the distance to the target from the origin, $\theta$ is the angular direction from the origin, and $\bm{u}(\theta) = [\cos \theta, \sin \theta]^T$ is the direction vector. We focus on a simple L\&S system with a single-antenna transmitter, allowing the L\&S to be described using a unified model. Let $x_{\text{pb}}(t)$ represent the transmitted passband pilot signal at time $t \in \mathbb{R}$, modulated to a carrier frequency $f_c$. By considering the free-space propagation in the radiating field and omitting the mutual coupling, the passband signal received by the $n$-th antenna from the target can be modeled as follows:
\begin{equation} \label{passband_model}
        y_{\text{pb},n}(t) = \frac{\beta}{r_n} x_{\text{pb}} \left( t - \tau_n + \tau_o \right),
\end{equation} 
where $\beta$ denotes the channel gain at a reference distance of 1 meter, $\tau_n$ is the signal delay from the target to the $n$-th antenna, and $\tau_o$ is the sampling clock offset at the receiver (which can be used for synchronization). The channel gain $\beta$ can be influenced by several factors, including signal wavelength, target material, antenna polarization, and antenna gain.
The signal delay $\tau_n = r_n/c$ depends on the distance $r_n$ from the target to the $n$-th antenna and the speed of light $c$. As shown in Figure \ref{fig:model}(a), the distance $r_n$ in near-field systems must be calculated as the exact Euclidean distance \cite{10220205}:
\begin{equation} \label{distance_NF}
        r_n = \| \bm{s}_n - \bm{p} \| = \sqrt{ r^2 - 2 r \bm{s}_n^T \bm{u}(\theta) + \|\bm{s}_n\|^2 }.
\end{equation}          
In wireless systems, most processing is conducted at the baseband, making it beneficial to abstract the carrier frequency and develop a baseband equivalent of the signal model. The baseband equivalents of the transmit and receive signals, $y_n(t)$ and $x(t)$, can be defined via $y_{\text{pb},n}(t) = \Re \left\{ \sqrt{2} y_n(t) e^{j 2 \pi f_c t} \right\}$ and $x_{\text{pb}}(t) = \Re \left\{ \sqrt{2} x(t) e^{j 2 \pi f_c t} \right\}$, where $\Re \{\cdot\}$ represents the real part of a complex number. Including the thermal noise $w_n(t)$, the baseband equivalent model is
\begin{equation} \label{baseband_model}
        y_n(t) = \frac{\beta}{r_n} x(t - \tau_n + \tau_o) e^{-j 2 \pi f_c (\tau_n - \tau_o)} + w_n(t).
\end{equation}
In narrowband systems, it then holds approximately that
$x(t - \tau_n + \tau_o) \approx x(t- r/c + \tau_o)$ due to the slow variation of the baseband signal across the array aperture. Then, by combining the signals from all antennas into a vector $\bm{y}(t) = [y_1(t),\dots,y_N(t)]^T$ and defining $s(t) = \beta/r \cdot x(t- r/c + \tau_o) e^{-j2\pi f_c (r/c-\tau_o)}$ as the common equivalent source signal received at all antennas, we obtain the narrowband model for near-field L\&S:
\begin{equation} \label{narrowband_model}
        \bm{y}(t) = \bm{a}(\theta, r) s(t) + \bm{w}(t),
\end{equation}
where $\bm{w}(t)=[w_1(t),\dots,w_N(t)]^T$ and $\bm{a}(\theta, r)$ is the near-field array response vector
\vspace{-0.1cm}
\begin{equation}
\label{eq:near-field-ARV}
        \bm{a}(\theta, r) = \left[\frac{r}{r_1} e^{-j \frac{2 \pi}{\lambda} (r_1-r)},\dots, \frac{r}{r_N} e^{-j \frac{2 \pi}{\lambda} (r_N-r)} \right]^T
\end{equation} 
Note that we write the near-field array response vector as a function of the target's direction $\theta$ and distance $r$ due to the dependence of $r_n$ on them. The amplitude variations in the near-field array response vector are often omitted since $r/r_n \approx 1$ holds whenever $r$ is substantially larger than the array aperture length \cite{Ramezani2023a, 10220205}. The model in \eqref{narrowband_model} can be easily extended to multi-target cases. If there are $K$ targets, where $K < N$, the receive signal becomes 
\begin{equation} \label{multi-target-signal}
    \bm{y}(t) = \bm{A}(\boldsymbol{\theta}, \bm{r}) \bm{s}(t) + \bm{w}(t),
\end{equation}
where $\bm{A}(\boldsymbol{\theta}, \bm{r}) = [ \bm{a}(\theta_1, r_1),\ldots,  \bm{a}(\theta_K, r_K)]$ and $\bm{s}(t) = [s_1(t),\ldots,s_K(t)]^T$ denote the array response matrix and equivalent source signals of $K$ targets, where $K < N$. The noise $\bm{w}(t)$ is generally assumed to be a white circular-symmetric complex Gaussian random process with a covariance matrix $\sigma^2 \bm{I}_N$. It is typically assumed that the array is calibrated and that the array response matrix is unambiguous, meaning that the matrix $\bm{A}(\boldsymbol{\theta}, \bm{r})$ has a rank of $K$, ensuring the unique identifiability of targets \cite{10144718}. To achieve this, it is widely assumed and accepted that the spatial version of Shannon’s sampling theorem needs to be satisfied, which requires the antenna separation to not be larger than $\lambda/2$. However, it is important to note that the requirements for unambiguous arrays can be different depending on the algorithm used. For instance, when both amplitude and phase variations of the near-field array response vector are exploited concurrently for L\&S, the $\lambda/2$ limit can be relaxed with particular processing \cite{9762062}, while for some low-complexity methods, a more stringent $\lambda/4$ limit is required \cite{He2012efficient}.

\begin{tcolorbox}[
title={\sc Near-field vs. far-field for fixed targets}, 
colback=white, colframe=black,
colbacktitle=white!80!gray, coltitle=black, 
breakable,
before upper={\parindent10pt\noindent},  
left = 1mm, 
right = 1mm,
top = 1mm,
bottom = 1mm
]
    \noindent In far-field L\&S, the target's distance is much larger than the array aperture. Therefore, the distance in \eqref{distance_NF} can be approximated as the following planar wave-based distance (see Figure \ref{fig:model}(b)):
    \begin{equation} \label{far_field_distance}
            r_n \approx r - \bm{s}_n^T \bm{u}(\theta).
    \end{equation}
    This approximation comes from the Taylor series $\sqrt{1 + x} = 1 + x/2 + \cdots$ for small $x = (-2r \bm{s}_n^T \bm{u}(\theta) + \|\bm{s}_n\|^2)/r^2$, ignoring higher order terms. Similar to \eqref{baseband_model}, we obtain the baseband far-field model as
    \begin{equation} \label{baseband_model_far}
            y_n^{\text{far}}(t) = \frac{\beta}{r} x(t - \tau_n^{\text{far}} + \tau_o)e^{-j 2\pi f_c (\tau_n^{\text{far}} - \tau_o)} + w_n(t),
    \end{equation} 
    where $\tau_n^{\text{far}} = (r-\bm{s}_n^T \bm{u}(\theta))/c$. Here, the channel gain $\beta/r$ is approximately the same at all antennas due to the large $r$.
    Inspecting \eqref{baseband_model} and \eqref{baseband_model_far}, we notice that both the far-field and near-field signals depend on the target's location ($\theta$ and $r$). This raises the question: What unique impacts does the near-field have on L\&S? To explore this, we need to identify which parameters are estimatable. A parameter is considered estimatable if it meets the consistency condition,
    meaning that there exists an estimator that can approximate the true value to any accuracy as the number of samples goes to infinity.  In wireless systems, samples can be collected across three dimensions: time, space, and frequency.
    Time-domain samples, acquired through sampling signals at different times $t$, provide an overall view of the noiseless received signal but fail to resolve $\theta$ and $r$, as these parameters are time-invariant. Therefore, the space- and frequency-domain samples are required. Applying the Fourier transform to the signals in \eqref{baseband_model} and \eqref{baseband_model_far} yields:
    \begin{align}
            Y_n(f) &= \frac{\beta}{r_n} X(f) e^{-j 2\pi \frac{(f_c + f) \sqrt{ r^2 - 2 r \bm{s}_n^T \bm{u}(\theta) + \|\bm{s}_n\|^2 } }{c}} + W_n(f), \\
            Y_n^{\text{far}}(f) &= \frac{\beta}{r} X(f) e^{ -j 2 \pi \frac{(f_c + f) r}{c} } e^{j 2 \pi \frac{(f_c + f) \bm{s}_n^T \bm{u}(\theta)}{c}} + W_n(f),
    \end{align}
    where the sampling offset $\tau_o$ is excluded for brevity.
    In the far-field model $Y_n^{\text{far}}(f)$, the observation of $r$ depends solely on the frequency $f$. Accurately estimating $r$ requires a wide bandwidth in $X(f)$. In a narrowband system, distance estimation is infeasible. However, the direction $\theta$ can be determined from space-domain samples taken at different antennas $\bm{s}_n$. Thus, the narrowband far-field signal is typically modeled as \cite{10144718}:
    \begin{equation}
            \bm{y}^{\text{far}}(t) = \bm{b}(\theta) s(t) + \bm{w}(t),
    \end{equation}
    where $\bm{b}(\theta) =[e^{j \frac{2\pi}{\lambda} \bm{s}_1^T \bm{u}(\theta)},\dots,e^{j \frac{2\pi}{\lambda} \bm{s}_N^T \bm{u}(\theta)} ]^T$ is the far-field array response vector (also known as steering vector). 
    The situation is different for the near-field signal $Y_n(f)$, for which the distance $r$ is related to both frequency $f$ and the antenna coordinate $\bm{s}_n$. Therefore, near-field provides a new way of estimating the target distance $r$ in the space domain, reducing the need for broad bandwidth.

\end{tcolorbox}

\subsection{Moving targets}
We now extend the model to a moving target within a millisecond-scale coherent processing interval (CPI), where the target approximately follows uniform motion along a straight line. Let $\bm{v} = v \bm{u}(\phi)$ denote the velocity vector of the target, where $v$ and $\phi$ represent the speed and direction, respectively. Within a CPI, the time-variant location of the moving target is $\bm{p}_o(t) = \bm{p} + \bm{v} t$. The distance from the target to the $n$-th antenna is thus given by 
\begin{equation}
        r_{o,n}(t) = \| \bm{s}_n - \bm{p}_o(t) \| = \sqrt{ r_n^2 - 2v t (\bm{s}_n - \bm{p})^T \bm{u}(\phi) + v^2 t^2 },
\end{equation}  
where $r_n$ is given in \eqref{distance_NF}. 
This expression is usually simplified using the Taylor series $\sqrt{1 + x} = 1 + x/2 + \cdots$ for small $x$.  Within a CPI, where $v t$ is almost always much less than the initial distance $r_n$, the distance $r_{o,n}(t)$ can be approximated as \cite{richards2005fundamentals}
\begin{equation} \label{moving_distance}
        r_{o,n}(t) \approx r_n - \frac{(\bm{s}_n - \bm{p})^T \bm{u}(\phi) v t}{r_n}. 
\end{equation}
In this formula, the term $(\bm{s}_n - \bm{p})^T \bm{u}(\phi)/r_n$ represents the projection from the velocity direction onto the line-of-sight direction of the $n$-th antenna. To highlight the difference from the far-field model, we redefine the velocity vector in terms of its components along and orthogonal to the line-of-sight path with respect to the origin: $\bm{v} = v_r \bm{u}(\theta) + v_{\theta} \bm{u}_\perp(\theta)$, where $\bm{u}_\perp(\theta) = [-\sin \theta, \cos \theta]^T$ is a unit vector orthogonal to $\bm{u}(\theta)$, and $v_r = v \bm{u}^T(\theta) \bm{u}(\phi)$ and $v_{\theta} = v \bm{u}_\perp^T(\theta) \bm{u}(\phi)$ denote the radial and transverse velocity, respectively. Substituting the new velocity vector into \eqref{moving_distance} yields 
\begin{equation} \label{moving_distance_new}
        r_{o,n}(t) = r_n + v_n t, \quad v_n = \underbrace{\frac{ (r - \bm{s}_n^T \bm{u}(\theta)) v_r}{r_n}}_{\text{Projection of $v_r$}} + \underbrace{\frac{-\bm{s}_n^T \bm{u}_\perp(\theta) v_{\theta}}{r_n}}_{\text{Projection of $v_{\theta}$}}.
\end{equation} 
Following \eqref{distance_NF} through \eqref{narrowband_model}, we obtain the narrowband near-field model for a moving target as \cite{wang2023velocity}
\begin{equation} \label{baseband_model_moving}
        \bm{y}_o(t) = \bm{a}(\theta, r) \odot \bm{d}(v_{\theta}, v_r, t)  s(t) + \bm{w}(t),
\end{equation}
where $\bm{d}(v_{\theta}, v_r, t)= [ e^{-j \frac{2\pi}{\lambda} v_1 t},\dots,e^{-j \frac{2\pi}{\lambda} v_N t} ]^T$ is the near-field Doppler vector at time $t$ and $\odot$ signifies the Hadamard product. For $K$ moving targets, the signal model becomes
\begin{equation} \label{multi-target-signal-moving}
    \bm{y}_o(t) = \left(\bm{A}(\boldsymbol{\theta}, \bm{r}) \odot \bm{D}(\bm{v}_{\theta}, \bm{v}_r, t)\right) \bm{s}(t) + \bm{w}(t),
\end{equation}
where $\bm{D}(\bm{v}_{\theta}, \bm{v}_r, t) = [\bm{d}(v_{\theta,1}, v_{r,1}, t), \ldots, \bm{d}(v_{\theta, K}, v_{r,K}, t)]$ denotes the Doppler matrix.

\begin{tcolorbox}[
title={\sc Near-field vs. far-field for moving targets},
breakable,
colback=white, colframe=black,
colbacktitle=white!80!gray, coltitle=black, 
before upper={\parindent10pt\noindent},  
left = 1mm, 
right = 1mm,
top = 1mm,
bottom = 1mm
]
    \noindent In far-field L\&S, where $r \gg D$, we have the following approximations are tight:
    \begin{equation}
            \frac{ r - \bm{s}_n^T \bm{u}(\theta)}{r_n} \approx 1, \quad \frac{-\bm{s}_n^T \bm{u}_\perp(\theta)}{r_n} \approx 0.
    \end{equation} 
    Combining this property with \eqref{far_field_distance}, the far-field time-variant distance can be approximated as
    \begin{equation}
            r_{o,n}(t) \approx r - \bm{s}_n^T \bm{u}(\theta) + v_r t.
    \end{equation}
    Thus, the narrowband far-field model for a moving target is
    \begin{equation} \label{baseband_model_moving_far}
            \bm{y}_o^{\text{far}}(t) = \bm{b}(\theta) e^{-j \frac{2\pi}{\lambda} v_r t} s(t) + \bm{w}(t). 
    \end{equation}
    Comparing \eqref{baseband_model_moving} and \eqref{baseband_model_moving_far}, we see that near-field effects cause a non-uniform Doppler frequency across the antenna array, providing additional information about the transverse velocity $v_{\theta}$. This allows us to determine the full motion state of the target in near-field L\&S. In contrast, far-field L\&S only provides information about the radial velocity $v_r$, making it challenging to ascertain the full motion state. 
\end{tcolorbox}

\section{Near-field L\&S for fixed targets}

The primary task for near-field L\&S with fixed targets is to estimate the location vectors $\boldsymbol{\theta}$ and $\bm{r}$ from the noisy signal in \eqref{multi-target-signal}, which is different from far-field L\&S where we can only estimate $\boldsymbol{\theta}$. However, the core nature of the estimation problem remains similar, allowing us to use conventional estimation techniques. Specifically, we aim to find the location vectors $\boldsymbol{\theta}$ and $\bm{r}$ that minimize or maximize an objective function based on a data matrix $\bm{X}$, derived from the observation matrix $\bm{Y} = [\bm{y}(1), \ldots, \bm{y}(L)] = \bm{A}(\boldsymbol{\theta}, \bm{r}) \bm{S} + \bm{W}$ where $\bm{S} = [\bm{s}(1),\ldots,\bm{s}(L)]$ and $\bm{W} = [\bm{w}(1),\ldots,\bm{w}(L)]$. In the following, we describe some of the well-known near-field L\&S methods, which estimate $\boldsymbol{\theta}$ and $\bm{r}$ either jointly or sequentially.

\subsection{Subspace fitting methods}

Many classical estimation methods, such as deterministic maximum likelihood (DML) and MUSIC, are specific instances of the subspace fitting method. These methods, widely used in far-field L\&S, can be extended to near-field cases. The general form of the subspace fitting problem for near-field L\&S is
\begin{equation} \label{subspace-fit}
    \{\hat{\boldsymbol{\theta}}, \hat{\bm{r}}\} = \argmin_{\boldsymbol{\theta}, \bm{r}} \min_{\bm{\Phi}} \| \bm{X} - \bm{A}(\boldsymbol{\theta}, \bm{r}) \bm{\Phi} \|_F^2,
\end{equation}
where $\bm{\Phi}$ is any unknown nuisance parameter and $\| \cdot \|_F$ is the Frobenius norm. This problem aims to fit the subspace spanned by $\bm{A}(\boldsymbol{\theta}, \bm{r})$ to the data matrix $\bm{X}$ in a least-squares manner. For fixed $\bm{A}(\boldsymbol{\theta}, \bm{r})$, the optimal nuisance parameter is $\hat{\bm{\Phi}} = \bm{A}^{\dagger}(\boldsymbol{\theta}, \bm{r}) \bm{X}$. Substituting this pseudoinverse solution back into \eqref{subspace-fit}, we get the equivalent problem
\begin{equation} \label{subspace-fit-con}
    \{\hat{\boldsymbol{\theta}}, \hat{\bm{r}}\} = \argmax_{\boldsymbol{\theta}, \bm{r}} \mathrm{tr} \left( \bm{P}_{\bm{A}}(\boldsymbol{\theta}, \bm{r}) \bm{X} \bm{X}^H \right),
\end{equation}
where $\bm{P}_{\bm{A}}(\boldsymbol{\theta}, \bm{r}) = \bm{A}( \bm{A}^H \bm{A})^{-1} \bm{A}^H $ is the projection matrix onto the column space of $\bm{A}(\boldsymbol{\theta}, \bm{r})$ and $\mathrm{tr}(\cdot)$ represents the trace of a matrix. The complex structure of $\bm{A}(\boldsymbol{\theta}, \bm{r})$ makes this problem highly non-convex, often resulting in many local maxima near the global maximum. Therefore, ensuring estimation accuracy typically requires a $2K$-dimensional high-resolution grid search over $\boldsymbol{\theta}$ and $\bm{r}$ for the region of interest, which can be computationally prohibitive in many application scenarios. Notice that a finer grid results in higher complexity but also greater estimation accuracy. However, infinite accuracy cannot be achieved by simply increasing grid granularity. This is because accuracy, which is usually evaluated by mean-squared error, is fundamentally limited by the theoretical Cramér–Rao bound (CRB), which will be discussed in the sequel.

The single-target approximation is a popular method to reduce the search space dimension \cite{10144718}. It estimates based on the single-target model while keeping the multi-target data matrix $\bm{X}$ unchanged, resulting in a simplified objective function: $\| \bm{X} - \bm{a}(\theta, r) \boldsymbol{\phi}^T \|_F^2$. Here, the multi-target array response matrix $\bm{A}(\boldsymbol{\theta}, \bm{r})$ is replaced by the single-target array response vector $\bm{a}(\theta, r)$. The locations of the $K$ targets are then found as the $K$ deepest local minima of this function. Substituting the optimal $\boldsymbol{\phi}$ with fixed $\bm{a}(\theta, r)$, the following equivalent problem can be obtained:
\begin{equation} \label{subspace_single_source}
    \{ \hat{\theta}, \hat{r} \} =  \text{ }^K \argmax_{\theta, r}  \mathrm{tr}\left(\bm{P}_{\bm{a}}(\theta, r) \bm{X} \bm{X}^H \right),
\end{equation}
where $\text{ }^K \argmax f(\cdot)$ identifies the $K$ arguments at which the function $f(\cdot)$ reaches its $K$ highest local maxima. While the single-target approximation reduces the dimension of grid-search from $2K$ to $2$, it may struggle with coherent signals from different targets. Such cases might require spatial smoothing to decorrelate the signals \cite{shan1985spatial}. The DML and MUSIC are the two most popular subspace fitting techniques that follow the general framework described above, as elaborated below.

\subsubsection{DML} 
In the DML method, the data matrix is directly chosen as the observation matrix, i.e., $\bm{X} = \bm{Y}$. The objective function in \eqref{subspace-fit} thus becomes $\| \bm{Y} - \bm{A}(\boldsymbol{\theta}, \bm{r}) \bm{S} \|_F^2$. Minimizing this objective function essentially maximizes the likelihood function of the observation signal $\bm{Y}$, assuming the source signal is deterministic and the noise is zero-mean complex circular Gaussian such that $\bm{y}(t) \sim \mathcal{CN}( \bm{A}(\boldsymbol{\theta}, \bm{r}) \bm{s}(t), \sigma^2 \bm{I}_N )$. The equivalent DML problem and its single-target approximation are given by
\begin{align}
    \{\hat{\boldsymbol{\theta}}, \hat{\bm{r}}\} &= \argmax_{\boldsymbol{\theta}, \bm{r}} \mathrm{tr} \left( \bm{P}_{\bm{A}}(\boldsymbol{\theta}, \bm{r}) \bm{Y} \bm{Y}^H \right), \\
    \label{SD_DML}
    \{ \hat{\theta}, \hat{r} \} &= \text{ }^K \argmax_{\theta, r}  \mathrm{tr} \left(\bm{P}_{\bm{a}}(\theta, r) \bm{Y} \bm{Y}^H \right).
\end{align}
From a computational complexity perspective, except for the initial calculation of the matrix $\bm{Y} \bm{Y}^H$, the overall complexity of the grid search for the DML method and its single-target approximation are $\mathcal{O}(N^3 (N_{g,\theta} N_{g,r})^K)$ and $\mathcal{O}(N^2 N_{g,\theta} N_{g,r})$, respectively. Here, $\mathcal{O}(\cdot)$ represents the big-O notation, and $N_{g,\theta}$ and $N_{g,r}$ denote the number of samples for direction and distance in the grid search, respectively.

\subsubsection{MUSIC}
In the MUSIC method, the data matrix $\bm{X}$ is derived from the covariance matrix of the observation signal $\bm{y}(t)$, theoretically given by $\bm{R} = \mathbb{E}[\bm{y}(t) \bm{y}^H(t)] = \bm{A} \bm{R}_s \bm{A}^H + \sigma^2 \bm{I}_N$, where $\bm{R}_s = \mathbb{E}[\bm{s}(t) \bm{s}^H(t)]$ and $\mathbb{E}[\cdot]$ denotes the statistical expectation. The rank of the signal subspace matrix $\bm{A} \bm{R}_s \bm{A}^H$ is at most $K$ when there are $K$ targets. In practice, we can estimate the covariance matrix as the sample average $\hat{\bm{R}} = \bm{Y} \bm{Y}^H/L$. The signal and noise subspaces are then estimated from the eigendecomposition $\hat{\bm{R}} = \bm{U}_s \bm{\Sigma}_s \bm{U}_s^H + \bm{U}_n \bm{\Sigma}_n \bm{U}_n^H$, where $\bm{\Sigma}_s$ and $\bm{\Sigma}_n$ contain the $K$ largest and $(N-K)$ smallest eigenvalues of $\hat{\bm{R}}$, and $\bm{U}_s$ and $\bm{U}_n$ contain the eigenvectors corresponding to the signal and noise subspaces, respectively. If $K$ is unknown, it can be estimated by identifying how many eigenvalues are substantially larger than the noise floor.

Following the subspace fitting concept, the data matrix can be selected as $\bm{X} = \bm{U}_s$ and the objective function can be designed as $\| \bm{U}_s -  \bm{A}(\boldsymbol{\theta}, \bm{r}) \bm{\Gamma} \|_F^2$, where $\bm{\Gamma}$ is a nuisance matrix related to $\bm{R}_s$. The corresponding equivalent problem and its single-target approximation are thus given by 
\begin{align}
    \label{MD_MUSIC}
    \{\hat{\boldsymbol{\theta}}, \hat{\bm{r}}\} &= \argmax_{\boldsymbol{\theta}, \bm{r}} \mathrm{tr} \left( \bm{P}_{\bm{A}}(\boldsymbol{\theta}, \bm{r}) \bm{U}_s \bm{U}_s^H \right), \\
    \label{SD_MUSIC}
    \{ \hat{\theta}, \hat{r} \} &=  \text{ }^K\argmax_{\theta, r} \mathrm{tr}\left(\bm{P}_{\bm{a}}(\theta, r) \bm{U}_s \bm{U}_s^H \right).    
\end{align}
The methods described in \eqref{MD_MUSIC} and \eqref{SD_MUSIC} are variants of the classical MUSIC method based on signal subspace \cite{viberg1991sensor}. It is easy to transform \eqref{SD_MUSIC} into the classical MUSIC based on the noise subspace \cite{Schmidt1986multiple}, \textcolor{blue}{\cite{Huang1991Near}}. Specifically, let $p(\theta, r) = \mathrm{tr}\left(\bm{P}_{\bm{a}}(\theta, r) \bm{U}_s \bm{U}_s^H \right)$, which satisfies $0< p(\theta, r) < 1$. Therefore, maximizing $p(\theta, r)$ is equivalent to maximizing $1/(1-p(\theta, r))$, resulting in the following classical MUSIC method:
\begin{equation}
        \label{C_MUSIC}
        \{ \hat{\theta}, \hat{r} \} = \text{ }^K \argmax_{\theta, r} \hspace{0.1cm} \frac{\|\bm{a}(\theta, r)\|^2}{\bm{a}^H(\theta, r) \left( \bm{I}_N - \bm{U}_s \bm{U}_s^H \right) \bm{a}(\theta, r)}. 
\end{equation}
The value of the numerator $\|\bm{a}(\theta, r)\|^2 = \sum_{n=1}^N (r/r_n)^2$ is approximately constant if the target is much further from the antenna array than the aperture length, so that $r/r_n \approx 1$. Using the orthogonality between the signal and noise subspaces, i.e., $\bm{I}_N - \bm{U}_s \bm{U}_s^H = \bm{U}_n \bm{U}_n^H$, we can simplify \eqref{C_MUSIC} as 
\begin{equation}
        \label{C_MUSIC_2}
        \{ \hat{\theta}, \hat{r} \} = \text{ }^K \argmax_{\theta, r} \frac{1}{\bm{a}^H(\theta, r) \bm{U}_n \bm{U}_n^H \bm{a}(\theta, r)}. 
\end{equation}
Similar to the DML methods, the computational complexity of the MUSIC method and its single-target approximation for grid search are $\mathcal{O}(N^3 (N_{g,\theta} N_{g,r})^K)$ and $\mathcal{O}(N^2 N_{g,\theta} N_{g,r})$, respectively. However, the MUSIC method also requires an additional eigendecomposition step, which has a complexity of $\mathcal{O}(N^3)$.

\begin{figure}
	\centering
	\begin{subfigure}{0.45\columnwidth}
		\centering
		\includegraphics[width=1\columnwidth]{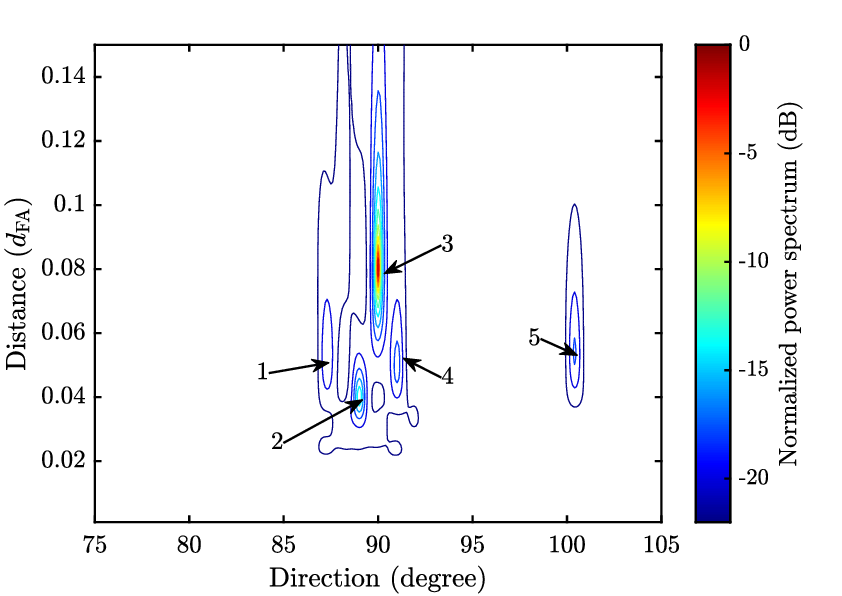}
		\caption{DML}
	\end{subfigure}
	\begin{subfigure}{0.45\columnwidth}
		\centering
		\includegraphics[width=1\columnwidth]{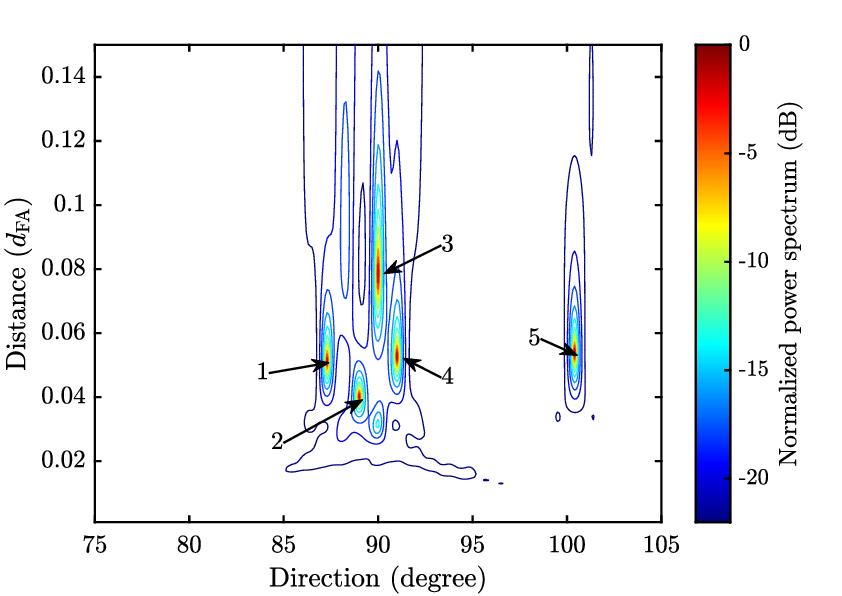}
		\caption{MUSIC}
	\end{subfigure}
 \caption{An example of the two-dimensional DML and MUSIC spectra obtained using the single-target approximation for five targets, with $28$ GHz carrier frequency, a uniform linear array (ULA) composed of $N = 128$ antennas, $L = 200$ data samples, and $\text{SNR} = 0$ dB. The ULA antenna spacing is set to $\lambda/2$. The MUSIC spectrum exhibits clearer peaks around the true target locations compared to the DML spectrum, particularly for closely located targets.}
 \label{fig:MUSIC_DML_Spectrum}
\end{figure}

While the DML method generally has superior asymptotic properties compared to MUSIC \cite{viberg1991sensor}, the latter performs better under single-target approximation, as illustrated in Figure \ref{fig:MUSIC_DML_Spectrum}. This improved performance of the MUSIC method is due to its effective utilization of the orthogonality between the signal and noise subspaces, which helps mitigate the interference between different targets. 
Due to the favorable properties, many MUSIC variants have been proposed in the literature for far-field L\&S to either reduce complexity or enhance estimation performance. Some variants rely on far-field properties, like the linear phase of the far-field array response used in the root-MUSIC method \cite{rao1989performance}, making them unsuitable for near-field cases. However, other variants, such as weighted MUSIC \cite{viberg1991sensor}, which do not depend on specific array response properties, remain effective for enhancing near-field L\&S performance.

\vspace{-0.2cm}
\subsection{Array-structure-based methods}

The structural information of the antenna array is important prior information that can be exploited to simplify the estimation problem and reduce computational complexity. For instance, the ESPRIT method, a well-known far-field L\&S technique designed for shift-invariant arrays, can eliminate the need for a grid search. Furthermore, in near-field L\&S, the DML and MUSIC methods incur a tremendous complexity, even with the single-target approximation, due to the two-dimensional search required over all possible directions and distances within the region of interest. Several methods have been proposed to address this issue by leveraging the symmetric structure of the array to turn the two-dimensional search into several one-dimensional searches, thereby remarkably reducing the complexity. Examples include the modified MUSIC method in \cite{He2012efficient} and the generalized ESPRIT method in \cite{Zhi2007nearfield}. This subsection focuses on the modified MUSIC algorithm which decouples the estimation of $\theta$ and $r$ by using the second-order statistics of the received signal \cite{He2012efficient}.

Assume the L\&S node has $N = 2M+1$ antennas arranged in a symmetric ULA, with the center antenna at the origin. The coordinate of the $n$-th antenna is $\bm{s}_n = [\delta_nd, 0]^T$, where $d$ is the inter-antenna spacing and $\delta_n = n - M -1$. The application of the symmetric ULA structure in the MUSIC method relies on two approximations: the amplitude variation of the near-field array response is negligible ($r/r_n \approx 1$), and the distance $r_n$ is approximated using the so-called Fresnel approximation as \cite{10220205}
\begin{equation}
    \label{Frensel_approx}
  r_n = \sqrt{r^2\left(1 - \frac{2\bm{s}_n^T \bm{u}(\theta)}{r} + \frac{\|\bm{s}_n\|^2}{r^2}\right)}  \approx r - \delta_n d \cos\theta + \frac{\delta_n^2 d^2 \sin^2\theta}{2r}, 
\end{equation}
which is obtained using the second-order Taylor approximation $\sqrt{1+x} = 1 + x/2 - x^2/8+ \cdots$ and ignoring some higher-order terms. Setting $\gamma = - 2\pi d \cos\theta / \lambda$ and $\psi = \pi d^2 \sin^2\theta / (\lambda r)$, the near-field array response vector can be reformulated from \eqref{eq:near-field-ARV} and \eqref{Frensel_approx} as
\begin{equation}
\label{eq:near-field-ARV_symmetric}
        \bm{a}(\theta,r) = \left[e^{j\left(-M\gamma + M^2 \psi\right) },\ldots,1,\ldots,e^{j\left(M\gamma + M^2 \psi\right) } \right]^T.  
\end{equation} 
Writing the array response vector in this form allows for decomposing the location estimation problem into separate direction and distance estimation problems. 
Specifically, with this symmetric structure of the near-field array response vector, the terms related to $\psi$ that involve the distance parameters cancel out in the anti-diagonal elements of the covariance matrix of the observation signal, i.e., $\bm{R} = \mathbb{E}\{\bm{y}(t) \bm{y}^H(t)\}$. Collecting all the anti-diagonal entries of $\bm{R}$ in a vector and defining $\epsilon_k = \mathbb{E}[s_k(t)s_k^*(t)]$, we have
\begin{equation}
\label{eq:anti_diagonal_vector}
 \bar{\bm{y}} = \sum_{k=1}^K \left[ \epsilon_k e^{-jM\gamma_k},\ldots, \epsilon_k e^{jM\gamma_k} \right]^T, 
\end{equation}
where we temporarily ignore the noise term for brevity. The elements in $\bar{\bm{y}}$ are independent of the target distances and only contain their direction information. This enables us to decouple the direction and distance estimation problems. 
To effectively estimate the $K$ directions via the MUSIC method, we need a covariance matrix of rank greater than $K$. This is to construct the signal and noise subspaces of sufficient dimensions.  To this end, the vector $\bar{\bm{y}}$ in \eqref{eq:anti_diagonal_vector} is divided into  $J$ overlapping subvectors, each having $N_s = 2M+2 - J$ entries. The $i$-th subvector is
\begin{equation} \label{m_MUSIC_subvector}
  \tilde{\bm{y}}(i) = \sum_{k=1}^K \left[ \epsilon_k e^{-j2(M-i+1) \gamma_k},\ldots, \epsilon_k e^{-j2(J-i-M) \gamma_k} \right]^T = \tilde{\bm{A}}(\boldsymbol{\gamma}) \tilde{\bm{s}}(i), 
\end{equation}
where $\tilde{\bm{s}}(i) = \left[\epsilon_1e^{j2i \gamma_1},\ldots,\epsilon_Ke^{j2i\gamma_K} \right]^T$ and $\tilde{\bm{A}}(\boldsymbol{\gamma}) = [\tilde{\bm{a}}(\gamma_1),\ldots,\tilde{\bm{a}}(\gamma_K)]$ with 
\begin{equation}
\label{eq:array_response_b}
    \tilde{\bm{a}}(\gamma_k) = \left[e^{-j2(M+1)\gamma_k},\ldots,e^{-j2(J-M)\gamma_k} \right]^T.
\end{equation}
The model in \eqref{m_MUSIC_subvector} exhibits a similar form to \eqref{multi-target-signal}. Therefore, the MUSIC method can be applied for estimating the $K$ directions (encapsulated in $\gamma_k$) by exploiting the signal and noise subspaces of the sample covariance matrix $\tilde{\bm{R}} =\sum_{i=1}^J \tilde{\bm{y}}(i)\tilde{\bm{y}}^H(i)/J$. Then, for each estimated direction, the corresponding distance can be found by employing the MUSIC method described in \eqref{C_MUSIC_2} but only need to search the distance dimension. This way, the two-dimensional grid search over $\theta$ and $r$ is replaced by $K+1$ one-dimensional grid searches. In particular, the computational complexity of the one-dimensional grid searches for $\theta$ and $r$ are $\mathcal{O}(N_s^2 N_{g,\theta})$ and $\mathcal{O}(N^2 N_{g,r})$, respectively, which are remarkably lower than the computational complexity of the conventional MUSIC method. However, this decoupling introduces a performance-complexity trade-off, as the modified MUSIC algorithm yields a sub-optimal estimate compared to the traditional approach.

\begin{figure}
	\centering
	\begin{subfigure}{0.45\columnwidth}
		\centering
		\includegraphics[width=1\columnwidth]{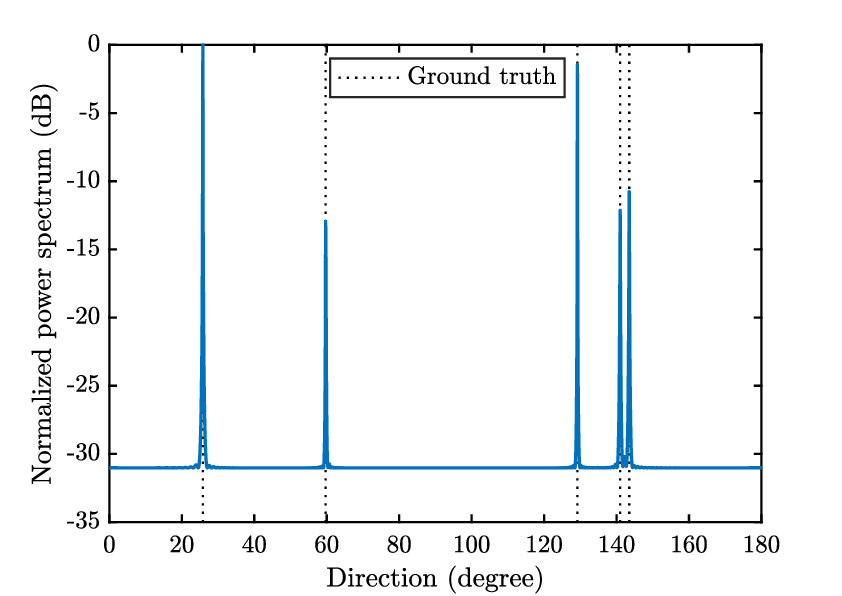}
		\caption{Direction spectrum}
	\end{subfigure}
	\begin{subfigure}{0.45\columnwidth}
		\centering
		\includegraphics[width=1\columnwidth]{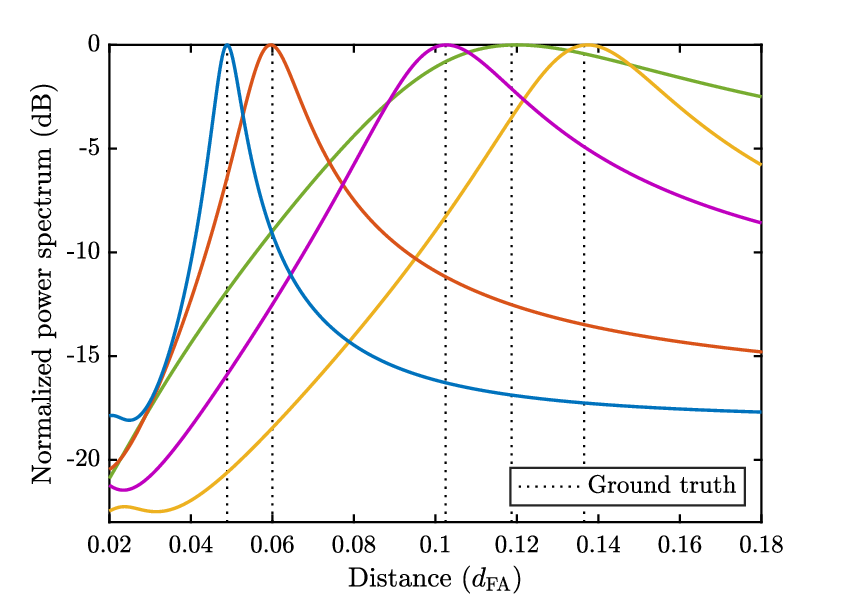}
		\caption{Distance spectrum}
	\end{subfigure}
 \caption{An example of the direction and distance spectra obtained using the modified MUSIC for five targets, with $N = 257$, $d = \lambda/4$, $J = 50$, $L=200$, and $\text{SNR} = 0$ dB. The carrier frequency is set to $28$ GHz. (a) The one-dimensional direction spectrum obtained by classical MUSIC method based on the model in \eqref{m_MUSIC_subvector}. (b) The five one-dimensional distance spectra obtained using the MUSIC method described in \eqref{C_MUSIC_2} at each estimated direction.}
 \label{fig:Modified_MUSIC}
 \vspace{-0.7cm}
\end{figure}

Note that there are three necessary conditions for the use of the modified MUSIC method: 1) a symmetric array structure to cancel out distance parameters in the anti-diagonal elements of the covariance matrix, 2) the Fresnel approximation in \eqref{Frensel_approx} is tight, typically requiring $r \ge 0.5 \sqrt{D^3/\lambda}$ \cite{10220205}, and 3) a small inter-antenna spacing $d \le \lambda/4$ to avoid direction ambiguity, c.f. \eqref{eq:array_response_b}. An example of the direction and distance spectra obtained using the modified MUSIC when the above conditions are met is depicted in Figure \ref{fig:Modified_MUSIC}. While the modified MUSIC algorithm reduces the computational complexity involved in estimating directions and distances, it may still require a significant number of searches, particularly when dealing with many targets. There are other array-structure-based methods that further simplify the estimation process by establishing a relationship between the estimation parameters, such as \cite{Zhang2018reduced} and \cite{Huang2024low}. These methods decompose the array response expression in \eqref{eq:near-field-ARV_symmetric} into a matrix-vector product with the matrix and vector only related to $\gamma$ and $\psi$, respectively, enabling closed-form estimators.

In this section, we have introduced the parametric subspace-based methods for near-field L\&S with fixed targets, like DML and MUSIC. These methods build on some common presumptions, such as the knowledge of the number of targets, low coherence between source signals, Gaussianity of signals and noise, and a large number of data samples, which limit their applicability to specific scenarios. The sparsity-based method, an important advancement inspired by compressive sensing in array signal processing, relaxes these assumptions by using a sparse signal model representation and is applicable to both near-field and far-field L\&S. However, it can be sensitive to hyperparameters and may bias estimation solutions. For more details on the sparsity-based method, see \cite{malioutov2005sparse}.

\vspace{-0.2cm}
\section{Near-field L\&S for moving targets}
\vspace{-0.05cm}
In a near-field L\&S system involving moving targets, it is crucial to account for both the targets' locations $\boldsymbol{\theta}$ and $\bm{r}$ and their velocities $\bm{v}_{\theta}$ and $\bm{v}_r$ encapsulated in the signal model in \eqref{multi-target-signal-moving}. This is a dynamical signal model due to the time-variant Doppler matrix $\bm{D}(\bm{v}_{\theta}, \bm{v}_r, t)$, which significantly complicates our estimation problem and renders many conventional methods inapplicable. This section will describe several potential estimation methods for near-field L\&S with moving targets.

\vspace{-0.25cm}
\subsection{Subspace fitting method}
The subspace fitting method remains useful for L\&S systems with moving targets and is widely used for estimating the direction $\theta$ and radial velocity $v_r$ in the far-field model \eqref{baseband_model_moving_far}.
In the far-field model, both DML and MUSIC can be effectively applied because all antennas experience the same Doppler frequency caused by $v_r$. For example, to estimate $\theta$ from \eqref{baseband_model_moving_far}, we can define $\tilde{s}(t) = e^{-j \frac{2\pi}{\lambda} v_r t} s(t)$ as an equivalent source signal, ensuring the signal subspace is always spanned by $\bm{b}(\theta)$ at different times. Thus, both DML and MUSIC can be used to estimate $\theta$. Moreover, $v_r$ can be estimated by DML, MUSIC, or Fourier analysis.

For the near-field model \eqref{multi-target-signal-moving}, the challenge is that different antennas observe different velocity components of targets, resulting in non-uniform Doppler frequencies and, thus, a time-varying signal subspace. This makes the covariance matrix and corresponding subspaces nearly immeasurable, rendering covariance matrix-based methods like MUSIC unusable. However, we can still apply the DML method. Specifically, the observation matrix over $L$, with a sampling interval of $T_s$, can be derived from \eqref{multi-target-signal-moving} as
$\bm{Y}_o = \bm{A}_o(\boldsymbol{\theta}, \bm{r}, \bm{v}_{\theta}, \bm{v}_r ) \tilde{\bm{S}} + \bm{W}$, 
where $\bm{A}_o(\boldsymbol{\theta}, \bm{r}, \bm{v}_{\theta}, \bm{v}_r ) = (\bm{1}_L^T \otimes \bm{A}(\boldsymbol{\theta}, \bm{r})) \odot \tilde{\bm{D}}(\bm{v}_{\theta}, \bm{v}_r)$ with $\tilde{\bm{D}}(\bm{v}_{\theta}, \bm{v}_r) = [\bm{D}(\bm{v}_{\theta}, \bm{v}_r, T_s), \ldots, \bm{D}(\bm{v}_{\theta}, \bm{v}_r, LT_s)  ]$ and $\tilde{\bm{S}} = \mathrm{blkdiag}\{\bm{s}(1),\ldots,\bm{s}(L)\}$. Here, $\otimes$ signifies the Kronecker product, $\mathrm{blkdiag}\{\cdot\}$ indicate block diagonal matrices, and $\bm{1}_L$ is an all-one vector of length $L$. Then, the DML method can be described as
\begin{equation}
    \{ \hat{\boldsymbol{\theta}}, \hat{\bm{r}}, \hat{\bm{v}}_{\theta}, \hat{\bm{v}}_r \} = \argmin_{\boldsymbol{\theta}, \bm{r}, \bm{v}_{\theta}, \bm{v}_r} \min_{\tilde{\bm{S}}} \| \bm{Y}_o - \bm{A}_o(\boldsymbol{\theta}, \bm{r}, \bm{v}_{\theta}, \bm{v}_r ) \tilde{\bm{S}}  \|_F^2.
\end{equation}
The computational cost of the above multi-dimensional DML is impractical for many applications due to the necessity of a $4K$-dimensional grid search. Additionally, calculating the projection matrix for the large-dimensional matrix $\bm{A}_o(\boldsymbol{\theta}, \bm{r}, \bm{v}_{\theta}, \bm{v}_r )$ presents further challenges. While reducing the DML search space dimension using single-target approximation is possible, it often leads to significant performance loss due to interference between targets. Partial relaxation offers a promising alternative by considering signals from all targets while maintaining a low-dimensional search space \cite{trinh2018partial}. However, extending this method to models with time-variant signal subspaces remains an open problem.

\vspace{-0.2cm}
\subsection{Nonlinear filtering method: from estimation to tracking}

As mentioned in the previous section, estimating target locations and velocities within a single CPI is challenging due to the lack of statistical information and the non-uniform Doppler frequencies caused by transverse velocity effects on the array response. However, if we consider a series of CPIs instead of focusing on a single CPI, this feature can actually facilitate target estimation. Let us take a single-target case as an example, where $r^i$, $\theta^i$, $v_r^i$, and $v_{\theta}^i$ represent the target state in the $i$-th CPI and $T_c$ denotes the CPI length. The target state between successive CPIs is highly correlated and we assume it can be described by the following state evolution model:
\begin{equation} \label{state_envolution}
    r^{i+1} = r^{i} + v_r^{i} T_c + z_r^i, \quad \theta^{i+1} = \theta^{i} + \frac{v_{\theta}^{i} T_c}{r^i} + z_{\theta}^i, \quad v_r^{i+1} = v_r^i + z_{v_r}^i, \quad v_{\theta}^{i+1} = v_{\theta}^i + z_{v_\theta}^i,
\end{equation}
where $z_r^i$, $z_{\theta}^i$, $z_{v_r}^i$, and $z_{v_\theta}^i$ are error parameters accounting for the perturbations caused by accelerations. The imposition of the above correlation constraint allows us to estimate the target state using both current CPI observations and previous CPI estimates, turning the estimation into a tracking problem over multiple CPIs. Note that the state evolution model \eqref{state_envolution} applies to both near-field and far-field scenarios. However, in the far-field, the absence of transverse velocity $v_{\theta}$ information in the observation signal makes it challenging to track the target accurately based on real-world measurements. 

Nonlinear filtering is a powerful tool to estimate and track the state of a nonlinear system from noisy observations. The extended Kalman filter (EKF) is the standard technique typically applied to a nonlinear system under additive Gaussian noise. We refer to \cite[Chapter 13.7]{kay1993fundamentals} for a detailed EKF derivation. The EKF requires a state evolution model and an observation model. The state evolution model is given in \eqref{state_envolution}, which can be written into a nonlinear function as 
\begin{equation}
    \bm{q}_{i+1} = h(\bm{q}_i) + \bm{z}_i, \quad \bm{z}_i \sim \mathcal{CN}(\bm{0}, \bm{R}_{\bm{z}}),
\end{equation}
where $\bm{q}_i = [r^i, \theta^i, v_r^i, v_{\theta}^i]^T$ and $\bm{z}_i = [z_r^i, z_{\theta}^i, z_{v_r}^i, z_{v_\theta}^i]^T$ are the state and noise vectors in the $i$-th CPI, respectively, $h(\cdot)$ is the nonlinear state evolution function, and $\mathcal{CN}(\bm{0}, \bm{C})$ denotes the circularly symmetric complex Gaussian distribution with covariance $\bm{C}$. The observation model is given in \eqref{baseband_model_moving}. In each CPI, numerous observation samples can be obtained by sampling the observation signal at different times. Exploiting more samples typically results in better tracking performance, especially for multi-target cases \cite{kamil2017multisource}. Despite this, it is still possible to use only one sample in each CPI. Without loss of generality, we consider the observation sample obtained at time $t = T_c/2$. Under this simplified condition, the observation model in the $i$-th CPI is given by
\begin{equation}
    \bm{y}_{o,i} =  g(\bm{q}_i) s_i + \bm{w}_i, \quad \bm{w}_i \sim \mathcal{CN}(\bm{0}, \bm{R}_{\bm{w}}),
\end{equation}
where $g(\bm{q}) = \bm{a}(\theta, r) \odot \bm{d}(v_{\theta}, v_r, T_c/2)$ is the nonlinear array response function and $s_i$ is the equivalent source signal at $t = T_c/2$ in the $i$-th CPI.  We are now prepared to formulate the EKF method, which can be summarized as the following iterative process:
\begin{enumerate}
    \item A priori state estimate: $\hat{\bm{q}}_{i|i-1} = h( \hat{\bm{q}}_{i-1} )$.
    \item Linearization: $\bm{H}_i = \left. \nabla_{\bm{q}} h(\bm{q}) \right|_{\bm{q} = \hat{\bm{q}}_{i-1}}$ and $\bm{G}_i = \left. \nabla_{\bm{q}} \left(g(\bm{q}) s_i \right) \right|_{\bm{q} = \hat{\bm{q}}_{i|i-1}}$.
    \item A priori covariance estimate: $\bm{Q}_{i|i-1} = \bm{H}_i \bm{Q}_{i-1} \bm{H}^H + \bm{R}_{\bm{z}}$.
    \item Kalman gain: $\bm{K}_i = \bm{Q}_{i|i-1} \bm{G}_{i}^H \left( \bm{G}_{i} \bm{Q}_{i|i-1} \bm{G}_{i}^H + \bm{R}_{\bm{w}}    \right)^{-1}$.
    \item A posteriori estimates: $\hat{\bm{q}}_{i} = \hat{\bm{q}}_{i | i-1} + \bm{K}_i \left( \bm{y}_{o,i} - g(\hat{\bm{q}}_{i | i-1}) s_i \right)$ and $\bm{Q}_i = \left( \bm{I} - \bm{K}_i \bm{G}_i   \right) \bm{Q}_{i|i-1}$.
\end{enumerate}

\begin{figure}
	\centering
	\begin{subfigure}{0.45\columnwidth}
		\centering
		\includegraphics[width=1\columnwidth]{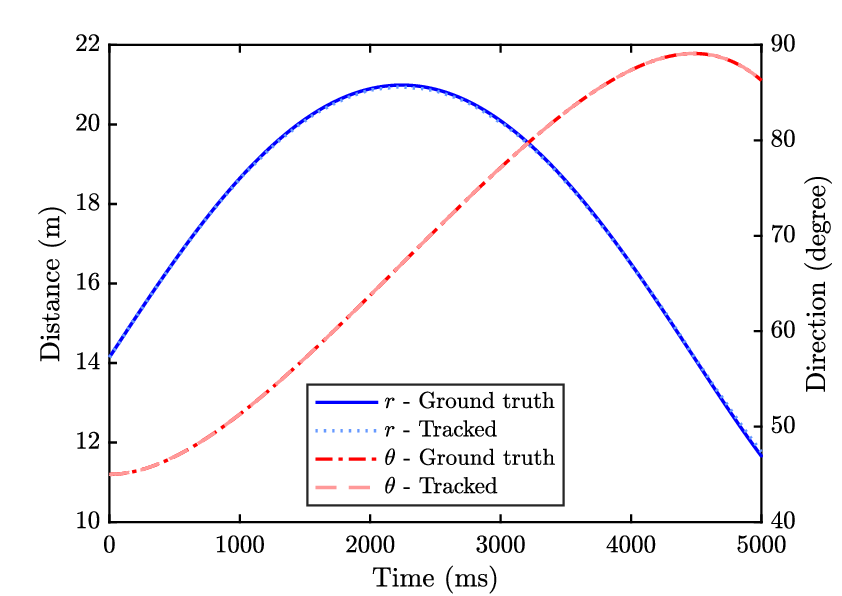}
		\caption{Location tracking}
	\end{subfigure}
	\begin{subfigure}{0.45\columnwidth}
		\centering
		\includegraphics[width=1\columnwidth]{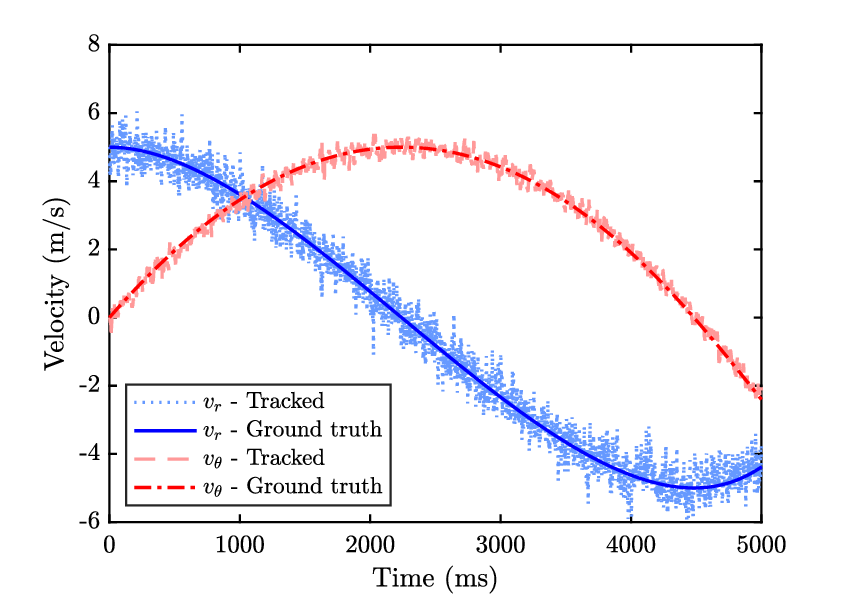}
		\caption{Velocity tracking}
	\end{subfigure}
 \caption{An example of applying the EKF to tracking a near-field moving target, with a $128$-antenna ULA, $28$ GHz carrier frequency, $T_c = 1$ ms, and $\text{SNR} = 0$ dB. The ULA antenna spacing is set to $\lambda/2$ and the mean-squared estimation error of $s_i$ is set to $-3$ dB. The covariance matrices of the state and observation noises are set to $\bm{R}_{\bm{z}} = \mathrm{diag}(0, 0, 0.01, 0.01)$ and $\bm{R}_{\bm{w}} = \sigma^2\bm{I}_N$, respectively.}
 \label{fig:EKF}
  \vspace{-0.7cm}
\end{figure}

Given that the dimension of the state vector $\mathbf{q}_i$ is much smaller than that of the observation vector $\mathbf{y}_{o,i}$, the computational cost of the EKF is primarily driven by the matrix inversion in Step 4, which has a worst-case complexity of $\mathcal{O}(N^3)$. When the number of $N$ becomes extremely large, such complexity may become unacceptable for real-time applications. One potential solution to this problem is to use the gradient descent method to avoid matrix inversion in Step 4. More specifically, it can be observed that the Kalman gain in Step 4 is essentially the solution to the following convex least squares problem:
\begin{equation}
    \min_{\bm{K}_i} \| \bm{K}_i \bm{V}_{i} - \bm{E}_{i} \|_F^2,
\end{equation}
where $\bm{V}_i = \bm{G}_{i} \bm{Q}_{i|i-1} \bm{G}_{i}^H + \bm{R}_{\bm{w}}$ and $\bm{E}_i = \bm{Q}_{i|i-1} \bm{G}_{i}^H$. In the field of machine learning, it has been well demonstrated that, compared to using the closed-form solution that involves matrix inversion, the gradient descent method can significantly reduce the complexity of solving the least squares problem when $N$ is extremely large. In recent years, leveraging neural networks to facilitate the real-time implementation of Kalman filters has also emerged as a promising solution \cite{9733186}. 

Furthermore, in Steps 2 and 5 of the EKF, knowledge of the equivalent source signal $s_i$ is required, which may be unavailable in practice. However, it can be estimated based on the a priori state estimate $\hat{\bm{q}}_{i|i-1}$ and the observation $\bm{y}_{o,i}$ \cite{kamil2017multisource}, which is essentially a signal detection problem under channel estimation error. An example of the EKF performance under imperfect knowledge of $s_i$ is shown in Figure \ref{fig:EKF}. Besides the robust tracking performance, another key advantage of the EKF method is its adaptability to multi-target scenarios. By following \eqref{state_envolution} and \eqref{multi-target-signal-moving}, we can obtain the multi-target state evolution model and the observation model, respectively. Then, the multi-target EKF can be carried out following the standard process. Apart from the EKF, other effective nonlinear filters based on Bayesian estimation are available in the literature \cite{haug2005tutorial}, including other variants of the Kalman filter, such as the unscented Kalman filter and the Monte Carlo Kalman filter, as well as the non-Gaussian filters, like particle filters, which can be helpful to improve the near-field L\&S performance in different scenarios.

\vspace{-0.2cm}
\section{Accuracy, resolution, and effective near-field region}

Accuracy and resolution are two critical performance metrics for L\&S systems. \emph{Accuracy} refers to the degree to which the estimated value of a parameter matches the true value. The CRB is an important indicator of accuracy, defined as the inverse of the Fisher information. It provides a theoretical lower bound on the mean-squared error of estimators for deterministic parameters. \emph{Resolution}, on the other hand, refers to the system's ability to distinguish between two or more closely spaced targets. In array processing, resolution is influenced by several factors, including array size, SNR, and the number of samples. The ambiguity function (AF), which represents the matched filter response, serves as a metric for assessing the array's ability to separate targets. Although the accuracy and resolution metrics serve different purposes, they are highly correlated: optimal accuracy is typically a scaling of the resolution by a factor related to the SNR. The accuracy and resolution for far-field L\&S have been extensively studied in the literature and can typically be represented by separate, simple, and closed-form CRBs and AFs. This simplicity stems from the fact that the key target parameters, such as distance, direction, and velocity, are not correlated in the far-field model.

\begin{figure}
\centering
 \includegraphics[width=0.45\columnwidth]{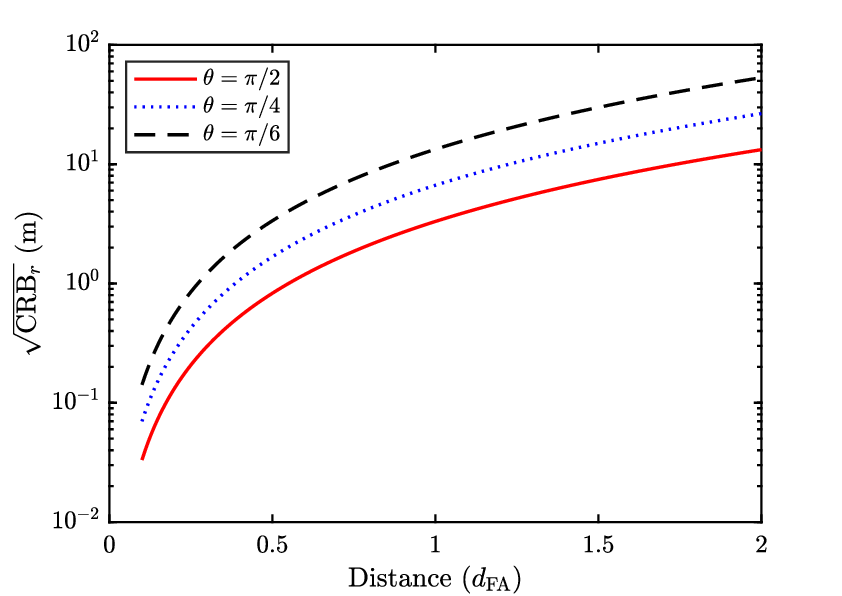}
 \caption{Numerical examples of CRBs for distance estimation with $N = 128$, $d = \lambda/2$, $L = 200$, and $\text{SNR}=0$ dB. The carrier frequency is set to $28$ GHz. The distance estimation exhibits the best performance when the target is positioned along the broadside of the antenna array, i.e., $\theta = \pi/2$.}
 \label{fig:CRB_distance}
  \vspace{-0.7cm}
\end{figure}

The situation is much more complicated in near-field systems because the target parameters become highly coupled and spatially variant, c.f. \eqref{narrowband_model} and \eqref{baseband_model_moving}. However, it is still possible to obtain relatively simple CRBs using the standard formula to evaluate the accuracy of near-field L\&S. In \cite{el2010conditional}, the closed-form CRBs for estimating $\theta$ and $r$ of a fixed target using a ULA with antenna spacing $d$ are derived by applying the Fresnel approximation in \eqref{Frensel_approx} and ignoring amplitude variation. These CRBs are given by
\begin{align}
    \label{CRB_theta}
    \mathrm{CRB}_{\theta} &= \frac{3 N_1}{2 \gamma L N_2 \pi^2 \sin^2 \theta }, \\
    \label{CRB_r}
    \mathrm{CRB}_r &= \frac{6 r^2 ( 15 r^2 + 30r d(N-1) \cos \theta + d^2N_1 \cos^2 \theta )}{\gamma L N_2 \pi^2 d^2 \sin^4 \theta },
\end{align}
where $\gamma = \text{SNR} \cdot d^2/\lambda^2$, $N_1 = (8N-11)(2N-1)$, and $N_2 = N(N^2-1)(N^2-4)$. Although the above CRBs are derived for a specific case, they provide qualitative insights that hold in general situations: 
\begin{itemize}
    \item The distance between the target and the L\&S node generally has a negligible impact on the accuracy of direction estimation (i.e., $\mathrm{CRB}_{\theta}$ is not related to $r$).
    \item The accuracy of distance estimation decreases unboundedly as the distance between the target and the L\&S node increases (i.e., $\mathrm{CRB}_r$ increases unboundedly with $r$, as shown in Figure \ref{fig:CRB_distance}.).
\end{itemize}
These insights align with intuition because as the distance approaches infinity, the near-field model converges to the far-field model, which only includes direction information. While the conclusions above were derived from the single-target CRBs, they generally hold true in multi-target scenarios as well. A recent study suggests a similar trend for the near-field velocity estimation with a moving target \cite{giovannetti2024performance}: a larger distance between the target and the L\&S node has negligible impact on the accuracy of radial velocity estimation but reduces the accuracy of transverse velocity estimation unboundedly. 

\begin{figure}
\centering
 \includegraphics[width=0.45\columnwidth]{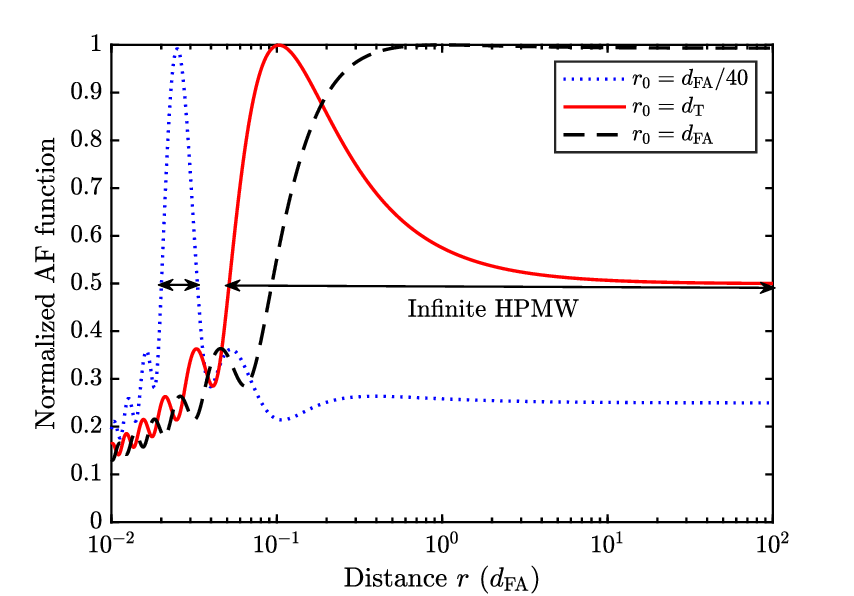}
 \caption{Illustration of the distance-domain AF $\Lambda_2(r)$ when $\theta_0 = \pi/2$.}
 \label{fig:AF_function}
  \vspace{-0.7cm}
\end{figure}

Regarding resolution, the AF can be characterized based on the matched filter response. Let's first consider the case of a fixed target located at $(\theta_0, r_0)$ so that the array response is time-invariant. Hence, we can represent the matched filter response of a single time-domain sample as the AF function, which is given by
\begin{equation}
    \mathrm{AF}( \theta, r ) = \mathbf{a}^H(\theta, r) \mathbf{a}(\theta_0, r_0).
\end{equation}
The half-power mainlobe width (HPMW), defined as the distance between the half-power intercepts of the AF mainlobe, serves as a baseline for determining how closely spaced two signals can be before they become challenging to resolve. Note that distinguishing two targets within the HPMW may require additional resources under noisy conditions, but it is not necessarily impossible. For instance, resolving two targets with overlapping HPMWs may be achievable with a significantly large number of data samples or a very high SNR, allowing subspace fitting methods like MUSIC to successfully separate them. A comprehensive analysis of the function $\mathrm{AF}( \theta, r )$ for unambiguous ULAs is provided in \cite{cui2022channel} from the codebook sampling perspective. It suggests that the direction resolution can be evaluated on a curve where $\sin^2\theta/r = \sin^2\theta_0/r_0$, leading to the AF function for the direction resolution
\begin{equation}
    \Lambda_1(\vartheta) \triangleq \mathrm{AF} \left( \theta, \frac{r_0 \sin^2 \theta}{\sin^2 \theta_0}  \right) \overset{(c)}{\approx} \frac{ \sin\left(  \pi N d (\vartheta - \vartheta_0)/\lambda \right)  }{\sin\left( \pi d (\vartheta - \vartheta_0)/ \lambda \right)},
\end{equation}
where $\vartheta = \cos\theta$ and $\vartheta_0 = \cos \theta_0$. The step $(c)$ uses the Fresnel approximation and omits the amplitude variation. $|\Lambda_1(\vartheta)|$ reaches its maximum value of $N$ at $\vartheta = \vartheta_0$. Then, the HPMW in the direction dimension is obtained as 
\begin{equation}
    |\Lambda_1(\vartheta_0 + \Delta \vartheta)|^2 \ge \frac{N^2}{2} \rightarrow \Delta \vartheta \in \left[-\frac{1.4 \lambda }{\pi N d},  \frac{1.4 \lambda }{\pi N d} \right].
\end{equation}
This is the same as the far-field resolution on the direction and is measured in radians, thus, the physical width is smaller at short distances. Moreover, the distance resolution can be evaluated by the following AF function:
\begin{equation}
    \Lambda_2(r) \triangleq \mathrm{AF}(\theta_0, r) \overset{(d)}{\approx} \frac{N C(\eta) + j N S(\eta) }{\eta},
\end{equation}
where $C(\eta) = \int_0^\eta \cos(\pi x^2/2) dx$ and $S(\eta) = \int_0^\eta \sin(\pi x^2/2) dx$ are Fresnel functions and $\eta$ is defined as 
\begin{equation}
    \eta = \sqrt{ \frac{N^2 d^2 \sin^2 \theta_0}{2\lambda} \left| \frac{1}{r} - \frac{1}{r_0} \right| }.
\end{equation}
The proof of $(d)$ can be found in \cite[Appendix A]{cui2022channel}. Similarly, $|\Lambda_2(r)|$ reaches its maximum value of $N$ at $r = r_0$. Then, the HPMW in the distance dimension is obtained as \cite[Appendix B]{10220205}
\begin{equation}
    \label{HPBW_distance}
    |\Lambda_2(r_0 + \Delta r)|^2 \ge \frac{N^2}{2} \rightarrow \Delta r \in \left[ - \frac{r_0^2}{d_{\mathrm{T}} + r_0}, \frac{r_0^2}{\max\{d_{\mathrm{T}} - r_0, 0\}} \right],
\end{equation}
where $d_{\mathrm{T}}$ is given by 
\begin{equation}
    \label{threshold_distance}
    d_{\mathrm{T}} \approx \frac{N^2 d^2 \sin^2 \theta_0}{5 \lambda}.
\end{equation}
An important observation from \eqref{HPBW_distance} is that the HPMW in the distance dimension is not always finite, as illustrated in Figure \ref{fig:AF_function}. Specifically, if $r_0 \ge d_{\mathrm{T}}$, the upper limit in \eqref{HPBW_distance} becomes $\infty$. The resolution of moving targets can be more complicated since the AF function must consider all time-domain samples in a CPI, which remains an open question. 

The above observations prompt us to reconsider the near-field region for L\&S. The classical Fraunhofer distance is defined based on the phase error from the far-field approximation relative to the accurate near-field model, which may not effectively reflect L\&S characteristics. Instead, we should reevaluate the effective near-field region by considering the accuracy and resolution of L\&S. Since the study of accuracy and resolution for moving targets is still in its early stages, we will focus on the fixed target case. For fixed targets, the key benefit of near-field L\&S is enabling distance estimation in narrowband systems. From this perspective, an important conclusion is that the effective near-field region can be considered infinite in terms of both accuracy and resolution. Specifically, for the distance accuracy evaluated by $\mathrm{CRB}_r$ in \eqref{CRB_r}, although $\mathrm{CRB}_r$ increases indefinitely with $r$, this increase can always be offset by higher SNR or more data samples $L$. As $\text{SNR} \to \infty$, $\mathrm{CRB}_r \to 0$, indicating infinite accuracy regardless of the distance from the L\&S node. A similar conclusion applies to resolution. In particular, for an unambiguous array, the ambiguity function has only a single maximum value regardless of distance, implying that two targets at different locations can always be distinguished in the noiseless case or as $\text{SNR} \to \infty$. However, in practice, a noiseless system is unattainable, and achieving very high SNR is challenging. In this context, the distance $d_{\mathrm{T}}$ in \eqref{threshold_distance}, beyond which distinguishing targets in the distance domain can be difficult in noisy environment, serves as a useful reference for defining the effective near-field region in practical systems. Importantly, $d_{\mathrm{T}}$ is much smaller than the Fraunhofer distance $d_{\mathrm{FA}}$. More specifically, in the case of $\theta_0 = \pi/2$ where $d_{\mathrm{T}}$ attains its maximum, we have 
\begin{equation}
    d_{\mathrm{T}} \approx \frac{N^2 d^2}{5 \lambda} \approx \frac{1}{10} d_{\mathrm{FA}},
\end{equation}
based on the expression of ULA's aperture $D = (N-1) d$. This result suggests that, from a distance resolution perspective, the effective near-field region is at most one-tenth of the Fraunhofer distance.

\vspace{-0.2cm}
\section{Conclusions and future directions}
In this article, we have provided a general review of the near-field L\&S techniques, from the fundamental signal models to basic signal processing methods for both fixed and moving targets. 
We have also discussed the accuracy and resolution from a theoretical perspective. 
We could not delve into many advanced methods for near-field L\&S due to space limitations, but we hope this article will inspire further literature reading, such as\cite{10144718, malioutov2005sparse, haug2005tutorial}. Despite the long history of near-field L\&S research, many challenges remain to be addressed. We highlight several promising directions for future research.

\begin{itemize}
    \item \emph{Performance bound characterization:} Most existing studies on the performance bounds of near-field L\&S are based on specific assumptions and approximations, such as uniform-power approximation and Fresnel approximation, often focusing on fixed target scenarios. As a result, these bounds may not hold in certain practical scenarios \cite{8736783}, such as extremely close targets, strong mutual coupling effects, or mobile targets. Therefore, further research is needed to characterize the performance of near-field L\&S based on an exact signal model, particularly for mobile targets.
    \item \emph{Wideband systems:} Although the near-field effect allows for estimating target distances in the spatial domain using narrowband signals, the achievable accuracy and resolution diminish as the target distance increases. As previously mentioned, using wideband signals, such as orthogonal frequency-division multiplexing (OFDM) and linear frequency modulation (LFM) signals, is an effective solution to counteract this diminishing accuracy. This approach introduces challenges such as large-dimensional space-frequency signals, spatial-wideband effects, and imperfect synchronization.
    \item \emph{Hybrid and low-resolution arrays:} Considering the large number of antennas and the high frequency in near-field systems, implementing digital antenna arrays presents significant challenges. Hybrid analog-digital arrays and low-resolution arrays are two popular alternatives to reduce hardware complexity and energy consumption \cite{bjornson2023twenty}. However, the use of these arrays can result in significantly distorted measurements or, in some cases, measurements with only one-bit resolution. Addressing such distortion for near-field L\&S remains an open problem.
    \item \emph{Sparse array design:} Sparse arrays, such as coprime and nested arrays, have been widely studied for far-field L\&S to improve direction estimation and increase the number of resolvable targets using virtual coarrays. In near-field L\&S, sparse arrays offer the additional benefit of generating a large aperture with fewer antennas, which improves the estimation performance of target distance and transverse velocity due to the more pronounced near-field effect. Tailored antenna placement optimization may be needed to fully reap the advantage of sparse arrays in near-field L\&S systems.
    \item \emph{Tensor-based signal processing:} Tensors are high-dimensional extensions of matrices. In near-field L\&S, the target states (distance, direction, radial velocity, transverse velocity) and data measurements (space, time, frequency) are multidimensional. Unlike far-field L\&S, where different target states can be estimated separately across different data dimensions, near-field L\&S involves highly coupled target states that require joint processing of the high-dimensional data. In this context, higher-order tensors can be a powerful tool \cite{7038247}.
    \item \emph{Data-driven and model-based machine learning:} The near-field L\&S methods discussed in this article are model-based and thus limited to specific scenarios. In particular, we have considered the free-space line-of-sight model, which is seldom exact in terrestrial use cases.
    In contrast, model-free data-driven approaches can generalize to various training data statistics but are typically data-hungry. In recent years, model-based machine learning has become a hot topic, which combines data and model knowledge to reduce data requirements and enhance robustness to model uncertainty and scattering and is thus promising to address the complicated near-field L\&S problems. Moreover, this method can also facilitate the implementation of real-world demonstrations of near-field L\&S systems using actual data.
\end{itemize}

\vspace{-0.35cm}
\bibliographystyle{IEEEtran}
\bibliography{mybib}

\end{document}